# Homochirality in an early peptide world

Axel Brandenburg[1,2], Harry J. Lehto[2,3] and Kirsi M. Lehto[2,4]

[1] Nordita, Blegdamsvej 17, DK-2100 Copenhagen Ø, Denmark

[2] Nordita, AlbaNova University Center, SE - 106 91 Stockholm, Sweden

[3] Tuorla Observatory and Physics Department, University of Turku, Finland

[4] Plant Physiology and Molecular Biology Laboratory, University of Turku, Finland

*Corresponding author*: brandenb@nordita.dk

**ABSTRACT**

A recently proposed model of non-autocatalytic reactions in dipeptide reactions leading to spontaneous symmetry breaking and homochirality is examined. The model is governed by activation, polymerization, epimerization and depolymerization of amino acids. Symmetry breaking is primarily a consequence of the fact that the rates of reactions involving homodimers and heterodimers are different, i.e., stereoselective, and on the fact that epimerization can only occur on the N-terminal residue and not on the C-terminal residue. This corresponds to an auto-inductive cyclic process that works only in one sense. It is argued that epimerization mimics both autocatalytic behavior as well as mutual antagonism - both of which were known to be crucial for producing full homochirality.

**INTRODUCTION**

Homochirality, i.e., the equal handedness of almost all amino acids (left-handed) and sugars (right-handed) is undoubtedly a striking property of all life on earth, and an essential requirement for the assembly of functional polymers (either polypeptides or nucleic acids). Indeed, the origin of homochirality is often thought to be closely associated with the origin of life itself (Avetisov 1991; Bada, 1995). Conversely, the amino acids of dead organisms gradually lose their preferred handedness. This is a well-known property of amino acids that is sometimes used as an approximate dating method (Hare and Mitterer 1967; Bada 1970). Since the chemistry of right and left-handed molecules is the same, it is conceivable that life could have been based on molecules whose handedness would be completely reversed. The selection of the two possible chiralities would then be a matter of chance, depending essentially on the presence of a minute initial excess of one over the other handedness. What is then required is a mechanism that amplifies any excess exponentially in time. If this were the case, it would not matter how small the initial excess was, provided the growth rate would be sufficiently large.

In a recent paper Plasson *et al*. (2004) attracted attention to the possibility that homochirality could have been attained in an early peptide world via a sequence of reactions producing dipeptides of only one handedness. They considered a closed, recycled system where the total number of building blocks is unchanged. In their model, monomers and dimers are coupled via activation, polymerization and depolymerization reactions. Here the activation is mediated via the formation of N-carboxyanhydrides.

Crucial ingredients in this model include the fact that the reaction rates for producing homodimers are different from those producing heterodimers (i.e., they are stereospecific), and the fact that epimerization occurs only on the N-terminal residue and not on the C-terminal residue. Their model carries the name APED, for activation, polymerization, epimerization, and depolymerization reactions. What is remarkable is the fact that apparently no autocatalysis is required, but the homochiralization process is based on what they call auto-induction. The preferential epimerization on the N-terminal residue is an empirically known fact (Kriausakul and Mitterer, 1980), although for some peptides preferential epimerization may also occur on the C-terminal residue (Kriausakul and Mitterer, 1983).

Since the seminal paper by Frank (1953) it has been considered that, on quite general grounds, two distinct ingredients are needed for establishing molecular symmetry breaking: autocatalysis and an inhibitory effect that Frank called mutual antagonism. Later, Sandars (2003) identified such an inhibitory effect as enantiomeric cross-inhibition in template directed polycondensation of polynucleotides (Joyce, 1984). However, autocatalytic reactions are not known to exist for small molecules such as short nucleic acids or short peptides. According to the APED model of Plasson *et al*. (2004) the stereoselective reactions favoring the formation of homochiral dipeptides, together with the coupled reaction network of polymerization, epimerization and depolymerization of amino acids, may produce an auto-inductive reaction cycle, leading to the same symmetry-breaking result as the classical hypothesis of an autocatalytic process with mutual antagonism.

The goal of the present work is to illuminate the similarity between the dipeptide reaction sequence proposed by Plasson *et al.* (2004) and the two governing effects of Frank's model, which are autocatalysis and mutual antagonism, and to investigate the effects of the reaction parameters of the original APED model, to illustrate its effects on the symmetry breaking.

**ESSENTIALS OF THE APED MODEL**

In their original paper Plasson *et al.* (2004) considered 8 pairs of reactions, including reactions for the activation and deactivation of both left and right-handed amino acids, spontaneous polymerization of activated amino acids into hetero- and homodimers, epimerization of the amino acids in the N-terminal position of the dimers, and depolymerization of the dimers. Reaction coefficients *a* and *b* were designated for activation and deactivation reactions, respectively, coefficients *p* and *h* respectively for polymerization and depolymerization of homodimers, and *e* for productive epimerization of the N-terminal amino acid to form homodimer; the non-stereospecific reaction rates involving corresponding homodimer reactions were quantified by reaction rates $\alpha p$, $\beta h$, and $\gamma e$. A pictorial overview of the set of the original set of reactions considered by Plasson *et al.* (2004) is given in Fig. 1.

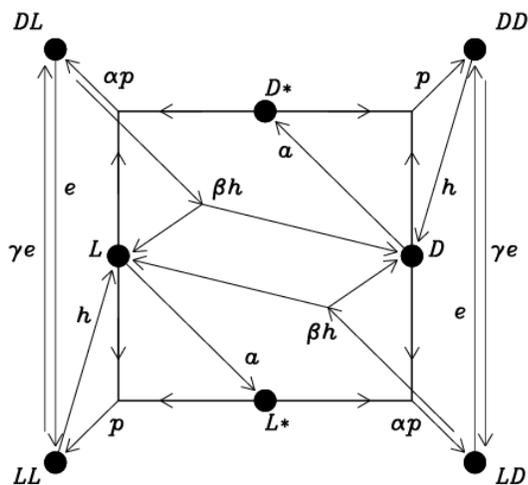

**FIG. 1.** Representation of the original set of reactions for producing homochirality.

The full set of all optional reactions is rather complex and hard to analyze, so Plasson *et al.* (2004) also considered an extreme and unrealistic case where the depolymerization and epimerization reactions were fully stereospecific ($\beta=\gamma=0$), and only the polymerization reaction varied between different degrees of stereospecificity (*αp*). Under these presumed conditions, 5 pairs of reactions determine the development of the symmetry state, and are adequate to obtain the remarkable effect of homochiralization in their model.

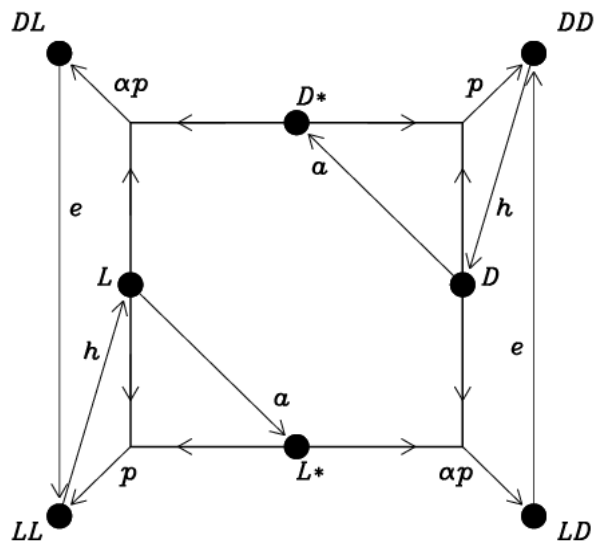

**FIG. 2.** Summary of essential and non-essential reactions of the APED model.

This minimal subset of reactions is shown in Fig. 2. It includes activation (proportional to the rate constant *a*), polymerization (proportional to the rate constant *p*), epimerization (proportional to the rate constant *e*), and depolymerization (proportional to the rate constant *h*). In addition, polymerization to produce heterodimers (proportional to the rate constant $\alpha p$) is critical for the APED mechanism to work. Unfortunately, even this minimal subset of reactions is still rather complex, so in order to understand what happens it is useful to consider meaningful limits in parameter space in which this subset of equations can be solved analytically, while still retaining the main mechanism of homochiralization. The resulting minimal set of equations necessary to retain this effect is summarized in the left-hand column of Box 1, while the remaining reactions are shown in the upper part of the right-hand column. In the lower part of the right-hand column we have also stated two additional reactions such as racemization ($L \leftrightarrow D$) with reaction rate *r*, and epimerization on the C-terminal position (Kriausakul and Mitterer, 1983) with reaction rates *g* and *ξg* for homochiral and heterochiral dimers, respectively.

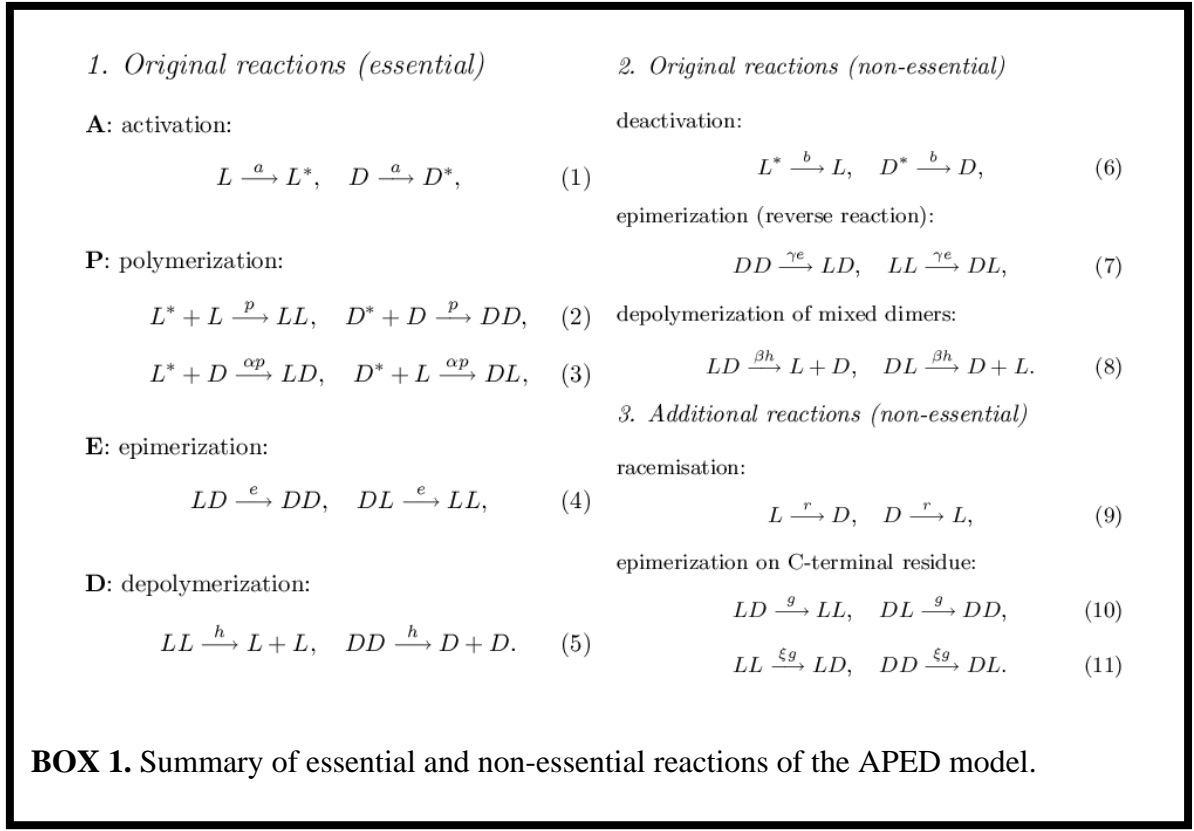

**BOX 1.** Summary of essential and non-essential reactions of the APED model.

Let us first illuminate the mechanism by which homochirality is achieved. This is best seen when writing the essential APED reactions in sequential form in one line, i.e.,

$$D^* + L \xrightarrow{\alpha p} DL \xrightarrow{e} LL \xrightarrow{h} L + L,$$
$$L^* + D \xrightarrow{\alpha p} LD \xrightarrow{e} DD \xrightarrow{h} D + D.$$

Thus, as long as the reaction rates for epimerization and depolymerization are not the limiting factors, we have essentially the reactions

$$D^* \xrightarrow{\alpha p[L]} L \quad \text{and} \quad L^* \xrightarrow{\alpha p[D]} D.$$

This way of writing these reactions emphasizes the roles of *L* and *D* in catalyzing the conversion of $D^*$ into *L* and $L^*$ into *D*, respectively. Plasson *et al.* (2004) introduced the term auto-induction instead of autocatalysis to emphasize the fact that autocatalysis in the normal sense is not thought to be possible with polymers as short as dimers. Thus, we can say that $L$ auto-induces the conversion of $D^*$ into $L$, and $D$ auto-induces the conversion of $L^*$ into $D$. In addition, there are reactions of the form

$$L^* + L \xrightarrow{p} LL \xrightarrow{h} L + L,$$
$$D^* + D \xrightarrow{p} DD \xrightarrow{h} D + D.$$

Again, these reactions simulate the autocatalytic conversion of $L^*$ into $L$ by $L$, and of $D^*$ into $D$ by $D$. These reactions, in the given conditions with fully stereoselective depolymerization and epimerization reactions can lead to full and sustained homochirality in situations where the value of $\alpha$ is between 0 and 1, i.e., the polymerization reaction is partially stereoselective (Plasson *et al.*, 2004).

In summary, the symmetry breaking described by the APED model seems to simulate autocatalytic behavior, even though the molecules themselves do not possess catalytic activity. In addition, the reactions involving the conversion from $D^*$ to $L$ and from $L^*$ to $D$ via stereoselective epimerization (if $\gamma=1$) reflect also mutual antagonism, but in an explicitly productive manner without producing achiral "waste" (degradation product). The chemical basis for these stereospecific reaction rates is not clear, but polymerization and epimerization reactions have been experimentally shown to behave in a stereospecific manner, favoring the formation of homodimers (Bartlett and Jones, 1957; Lundberg and Doty, 1956; Commeyras, 2002; Plasson, 2003), and could be caused for instance by stereochemical a stacking effect of the two amino acids. In order to put this on a more mathematical basis, we consider now the kinetic equations of a minimal subset of the APED model. For simplicity, deactivation and depolymerization of heterodimers, as well as epimerization to produce heterodimers, are neglected. This corresponds to the presumed initial setting $b=\beta=\gamma=0$ in Plasson *et al.* (2004). The resulting set of equations is given in Box 2.

$$\frac{d}{dt}[L] = -a[L] - p\big([L^*] + \alpha[D^*]\big)[L] + 2h[LL], \qquad \frac{d}{dt}[LL] = p[L^*][L] + e[DL] - h[LL],$$

$$\frac{d}{dt}[D] = -a[D] - p\big([D^*] + \alpha[L^*]\big)[D] + 2h[DD], \qquad \frac{d}{dt}[DD] = p[D^*][D] + e[LD] - h[DD],$$

$$\frac{d}{dt}[L^*] = a[L] - p\big([L] + \alpha[D]\big)[L^*], \qquad \frac{d}{dt}[DL] = \alpha p[D^*][L] - e[DL],$$

$$\frac{d}{dt}[D^*] = a[D] - p\big([D] + \alpha[L]\big)[D^*], \qquad \frac{d}{dt}[LD] = \alpha p[L^*][D] - e[LD].$$

**BOX 2.** Kinetic equations corresponding to the minimal subset of equations essential for the APED model to work.

Let us now use explicitly the assumption that epimerization and depolymerization are not the limiting factors in the reaction and that these two reactions are much faster than the activation step, i.e., both $e$ and $h$ are large compared with $a$. We consider this case mainly in order to illuminate the nature of the multi-step auto-inductive reaction displayed above. Thus, *DL* evolves rapidly via *LL* into *L+L*, and, *LD* evolves rapidly via *DD* into *D+D*. Mathematically, this is achieved by removing the time derivatives for the last four of the essential reactions equations. This reduces the number of explicitly time dependent equations to four. The resulting system of equations is written below.

$$\frac{d}{dt}[L] = -a[L] + p\big([L^*] + \alpha[D^*]\big)[L],$$

$$\frac{d}{dt}[D] = -a[D] + p\big([D^*] + \alpha[L^*]\big)[D],$$

$$\frac{d}{dt}[L^*] = a[L] - p\big([L] + \alpha[D]\big)[L^*],$$

$$\frac{d}{dt}[D^*] = a[D] - p\big([D] + \alpha[L]\big)[D^*].$$

**BOX 3.** Kinetic equations corresponding to the reduced set of equations containing the essentials of the APED model.

This new reduced system of equations permits a simple and interesting interpretation in that it too can be associated with chemical reactions:

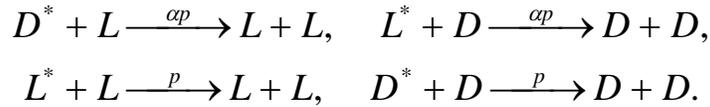

$$D^* + L \xrightarrow{\alpha p} L + L, \quad L^* + D \xrightarrow{\alpha p} D + D,$$
$$L^* + L \xrightarrow{p} L + L, \quad D^* + D \xrightarrow{p} D + D.$$

These reactions, together with the corresponding activation steps, have been depicted in Fig. 3, and they are indeed equivalent to the multi-step reactions discussed above. Qualitatively, the process can be explained as follows. A small initial excess of, say, [L] over [D] enhances the supply of [L] from [$D^*$], which appears to be like autocatalysis. The diminished level of [$D^*$] enhances the losses of [D] toward [$D^*$], because of the minus sign in the corresponding rate $a$-$p$[$D^*$]. The reduced level of [D] appears like "productive" mutual antagonism. This also decreases the losses of [$L^*$] toward [D], so [$L^*$] stays high and hence losses of [L] toward [$L^*$] are minimized, because of the minus sign in the corresponding reaction rate $a$-$p$[$L^*$].

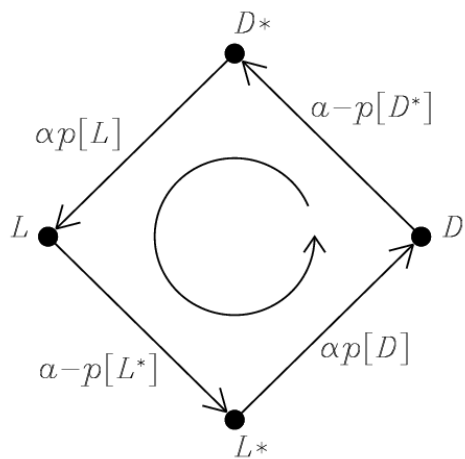

**FIG. 3.** Representation of the reduced set of reactions leading to homochirality. Note the counter-clockwise sense of the reaction sequence.

The equations can be reduced further to only two explicitly time-dependent equations if we assume that also $p$ is large ($cp \gg a$, where $c$ is the total concentration of all building blocks, which is a constant in this model). It turns out that the enantiomeric excess obeys the equation

$$\frac{d}{dt}(\text{e.e.}) = \lambda \times (\text{e.e.}),$$

where $\lambda$ is the growth rate, which is given by

$$\lambda = \frac{2a\alpha(1-\alpha)[L][D]}{(1+\alpha^2)[L][D] + \alpha([L]^2 + [D]^2)}.$$

For the racemic solution the growth rate of the instability toward homochirality is

$$\lambda = 2a\alpha \frac{1-\alpha}{(1+\alpha)^2}.$$

Evidently, and in agreement with Plasson *et al.* (2004), the growth rate is positive as long as $\alpha < \alpha_{\text{crit}}$, where $\alpha_{\text{crit}}=1$ is the maximum possible value for achieving still homochirality in the reduced model. Once one of the two homochiral states has been reached, either [D] or [L] vanish and hence $\lambda=0$, terminating further growth.

We see that the enantiomeric excess shows exponential growth whenever $\alpha$ is between 0 and 1. It is remarkable that this criterion is so general and apparently independent of the values of the other parameters. However, we ought to remember that several restrictive approximations have been made in arriving at this result, most notably that *h*, *e*, and *p* were assumed large, and β and γ were assumed $= 0$.

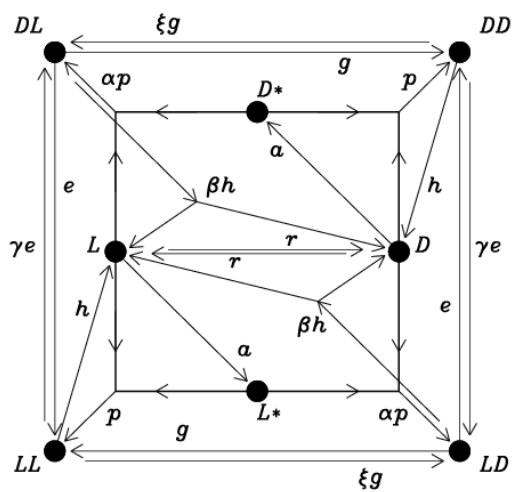

**FIG. 4.** Representation of the full and expanded set of reactions leading to homochirality.

In Fig. 4 we show an expanded set of reactions and demonstrate in Fig. 5 that even for smaller values of $h$, $e$, and $p$, and also for finite values of $\beta$ and $g$, the criterion is unchanged, and that only the values of the growth rates change. The only reaction that changes this criterion is the racemization reaction, characterized by the parameter $r$. If $r$ is larger than $0.12a$, only the racemic solution is possible; see Fig. 6, while for $r$ less than $0.12a$ there is a finite interval of $\alpha$ where a homochiral (right or left-handed) solution is possible. For $r=0.05$, example, homochirality is only possible when $\alpha$ is in the interval between 0.06 and 0.80; see Fig. 6.

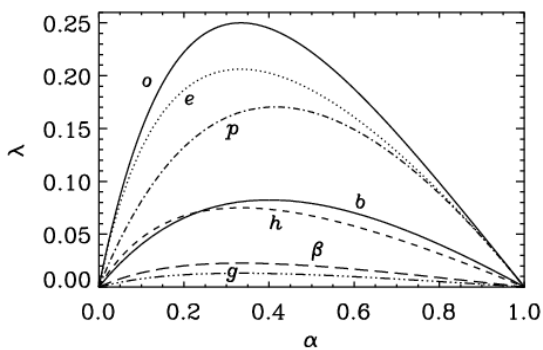

**FIG. 5.** Growth rate versus $\alpha$ for different parameter combinations. The solid line (denoted $o$) gives the asymptotic formula described in the text, while for all the other curves one parameter is different from a value that would reproduce the asymptotic result ($e=100a$, $p=200a/c$, $h=1000a$, $b=g=\beta=0$). For the dotted curve (denoted $e$) we have $e=a$, for the dash-dotted curve (denoted $p$) we have $p=2a/c$, for the dashed curve (denoted $h$) we have $h=a$, for the other solid line (denoted $b$) we have $b=100a$, for the long-dashed curve (denoted $\beta$) we have $\beta=1$ (!), and for the triple dot-dashed curve (denoted $g$) we have $g=0.9e$.

In conclusion, the interpretation of a one-way circular reaction scheme based on the simplified model presented above appears to be robust, and we may therefore conclude that the APED model does indeed capture effects quite analogous to the usual autocatalysis and mutual antagonism phenomena.

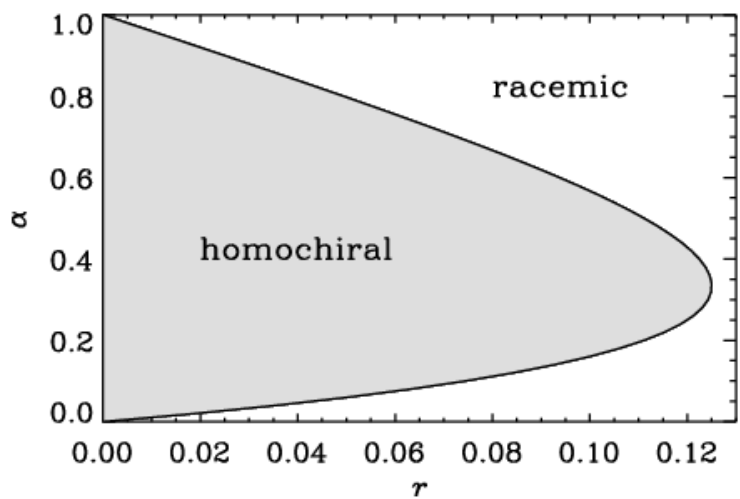

**FIG. 6.** Chirality regimes as a function of α and the racemization parameter *r*.

**TEMPORAL EVOLUTION**

If the initial condition were exactly racemic, homochirality would of course never emerge. However, such a special initial condition would be quite unrealistic, and there will always be a distribution of the initial value of e.e. around zero. The width of this distribution decreases with increasing number of molecules that can interact (the width is $1/\sqrt{n}$ for *n* molecule

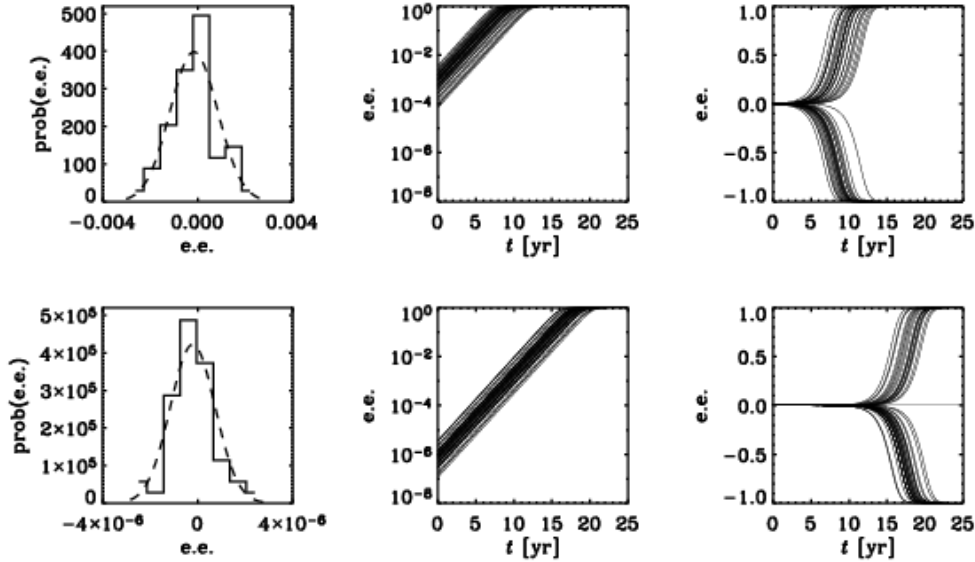

**FIG. 7.** Two examples of the probability distribution of the initial e.e. for racemic mixtures with $10^6$ and $10^{12}$ molecules together with the resulting evolution of e.e., both in logarithmic and linear representations, using $a=10^{-8}$ s$^{-1}$. The dashed lines give a gaussian fit to the distribution function.

We illustrate this in Fig. 7 by plotting the evolution of e.e. for two different random initial distributions of molecules. In addition to plotting the distributions of the initial values of e.e., we also show logarithmic and linear plots of the evolution of e.e., which shows quite clearly that after a time of about 7-14 times the value of $1/a=3$ yr (for $a=10^{-8}$ s$^{-1}$ quoted by Plasson *et al.*, 2004), full homochirality is achieved. This time depends only logarithmically on the initial e.e., so for e.e. $= 10^{-3} - 10^{-6}$ on has $\log(10^3 - 10^6) = 7 - 14$ times the value of $1/a$.

**FINAL REMARKS**

It is not clear under which circumstances the circle of reactions described above could have operated. Is it a phenomenon that might have occurred naturally on the early earth, and could a similar auto-inductive set or reactions have worked also on ribonucleotides? If so, homochirality may well have been an important condition that might have enabled the formation of sufficiently long ribonucleotide polymers and maybe, indirectly, the emergence of life. The work of Plasson *et al.* (2004) may well be interpreted as pointing in this direction. The other alternative would be that homochirality developed as a consequence of enantiomeric cross-inhibition combined with autocatalysis during a long "struggle" of short self-replicating RNA molecules for dominance, as envisaged by Sandars (2003) in his model; see also Brandenburg (2005). The difficulty here is that autocatalysis is required, which may be difficult with short nucleic acids.

One may well imagine a combination of an early peptide world providing a homochiral environment, together with a developing RNA world where sufficiently long isotactic autocatalytic molecules have been synthesized. Although autocatalysis may not have been operational in prebiotic chemistry, the catalysis on clay surfaces remains an interesting and frequently discussed possibility (Schwartz, 1996; Yu 2001; Cintas, 2002). Nevertheless, it seems likely that, if complete homochirality did emerge as a result of spontaneous symmetry breaking of any kind, the crucial ingredients would still be self-amplification and competition, much like in the original model of Frank (1953).